\documentclass[%
 reprint,
 amsmath,amssymb,
 prl,
]{revtex4-1}

\usepackage{graphicx}
\usepackage{dcolumn,hyperref}
\usepackage{bm}

\begin{document}


\title{On the origin of the LIGO ``mystery'' noise and the high energy particle physics desert}

\author{Niayesh Afshordi}%
 \email{nafshordi@pitp.ca}
\affiliation{Department of Physics and Astronomy, University of Waterloo, 200 University Ave W, N2L 3G1, Waterloo, Canada}
\affiliation{Waterloo Centre for Astrophysics, University of Waterloo, Waterloo, ON, N2L 3G1, Canada}
\affiliation{Perimeter Institute For Theoretical Physics, 31 Caroline St N, Waterloo, Canada}

\date{\today}

\begin{abstract}
One of the most ubiquitous features of quantum theories is the existence of zero-point fluctuations in their ground states. For massive quantum fields, these fluctuations decouple from infrared observables in ordinary field theories. However, there is no ``decoupling theorem'' in Quantum Gravity, and we recently showed that the vacuum stress fluctuations of massive quantum fields source a red spectrum of metric fluctuations given by $\sim$ mass$^5$/frequency in Planck units. I show that this signal is consistent with the reported unattributed persistent noise, or ``mystery'' noise, in the Laser Interferometer Gravitational-Wave Observatory (LIGO), for the Standard Model of Particle Physics. If this interpretation is correct, then it implies that: 1) This will be a fundamental irreducible noise for all gravitational wave interferometers, and 2) There is no fundamental weakly-coupled massive particle heavier than those in the Standard Model.       

\end{abstract}

\maketitle

Finding a unified framework for describing the quantum theory and gravity is one of the most outstanding puzzles in the foundations of natural sciences. It is often thought that a theory of ``Quantum Gravity'' is only necessary to understand Planck-scale physics, e.g.,  short distances $\lesssim 10^{-33}$ cm or high temperatures $\gtrsim 10^{32}$ K. However, the puzzles and paradoxes that arise when ``Quantum'' meets ``Gravity'' have surprising manifestations on much larger scales, ranging from the horizons of black holes (i.e. information paradox \cite{Almheiri:2012rt}) to the dark energy (i.e. cosmological constant problems \cite{Weinberg:1988cp,Weinberg:2000yb}).   

In lieu of a tractable ``UV-complete'' theory of Quantum Gravity (QG), Effective Field Theory (EFT) arguments are often used to quantify the QG effects in the IR. While EFT framework has been extremely successful in treating local quantum fields theories, its application to QG is suspect (e.g., \cite{Donoghue:2009mn,Afshordi:2015iza}): \begin{enumerate}
    \item Decoupling theorems (such as that of Appelquist \& Carrazone \cite{Appelquist:1974tg}) are often invoked for why heavy particles decouple from gauge fields in the IR. However, these theorems are only applicable to renormalizable gauge theories which precludes gravity. In other words, while virtual heavy particle/anti-particle pairs can screen each other due to opposite charges, effectively decoupling from light gauge fields, the ``gravitational'' charge is always positive and thus there is {\it no} ``gravitational screening''.   
\item In contrast to other gauge symmetries, the diffeomorphism symmetry of QG is nonlocal, as it maps widely-separated points in spacetime to each other. This makes a full theory of QG inherently nonlocal, and thus there is no reason it should reduce to a local EFT in the IR.  

\item Another key element for the consistency of EFT is the separation of energy scales: Heavy fields remain in their adiabatic ground state during low-energy processes, as there is not enough energy to excite them. However, there is no notion of energy conservation in QG if spacetime is not asymptotically flat (more on this later). 

\item A crucial ingredient for the usefulness of EFT is ``technical naturalness'', providing an organizing principle for the infinitely many interaction terms that arise in EFT \cite{tHooft:1979rat}. However, the fact that the cosmological constant has a technically unnatural small value \cite{Weinberg:1988cp,Weinberg:2000yb} brings into question the applicability of this organizing principle, and thus EFT, in QG.
\end{enumerate}

More speculative arguments against technical naturalness and UV-IR decoupling in QG, which are often invoked in the EFT framework, have been recently discussed under the title of ``swampland conjectures'' \cite{Vafa:2005ui,ArkaniHamed:2006dz,Obied:2018sgi}. 

With these motivations, we recently explored how the stress fluctuations of a massive quantum field would back-react on the gravitational vacuum in the IR \cite{Afshordi:2015iza,Afshordi:2017scc}. We discovered that, for free quantum fields of arbitrary spin, the symmetrized connected correlators of the stress tensor in Minkowski spacetime take a universal form if analytically continued to the IR \cite{Afshordi:2017scc}:
\begin{eqnarray}
&&\langle T_{\mu\nu}(x)T_{\alpha\beta}(y)\rangle_{\rm IR} = -\frac{1}{60\pi} \times \left(\sum_i c_i m_i^5\right)\times\nonumber\\ &&\int \frac{d^4k}{(2\pi)^4} \frac{\exp[i k\cdot (x-y)]}{\sqrt{k^2}} \left(P_{\mu\nu}P_{\alpha\beta} + P_{\mu\alpha}P_{\nu\beta} +P_{\mu\beta}P_{\nu\alpha}\right),  \nonumber\\ \label{stress}
\end{eqnarray}
 where $m_i$ and $c_i$ are the mass and the number of polarizations of the $i$'th field, respectively, while
 $P_{\mu\nu}   \equiv \eta_{\mu\nu}-k_\mu k_\nu/k^2$ is the projector tensor \footnote{We use $(-,+,+,+)$ signature}.
We then proceeded to present a covariant derivation of pulsar timing noise, if the stress fluctuations in Equation (\ref{stress}) were to source metric fluctuations. This yields a dimensionless strain noise of
\begin{equation}
\langle h^2_{\rm pulsar} \rangle = \frac{7 \sum_i c_i m_i^5}{240 \pi^3 M^4_{\rm P}} \sqrt{\rm distance \over 2 \times frequency},    
\end{equation}
where $M_{\rm P} = 2.44 \times 10^{18}$ GeV is the reduced Planck mass. In other words, deviations from flat spacetime will diverge over long times or large distances, what we dubbed the ``cosmological non-constant problem'' \cite{Afshordi:2015iza}.

\begin{figure}[b]
\includegraphics[width=\linewidth]{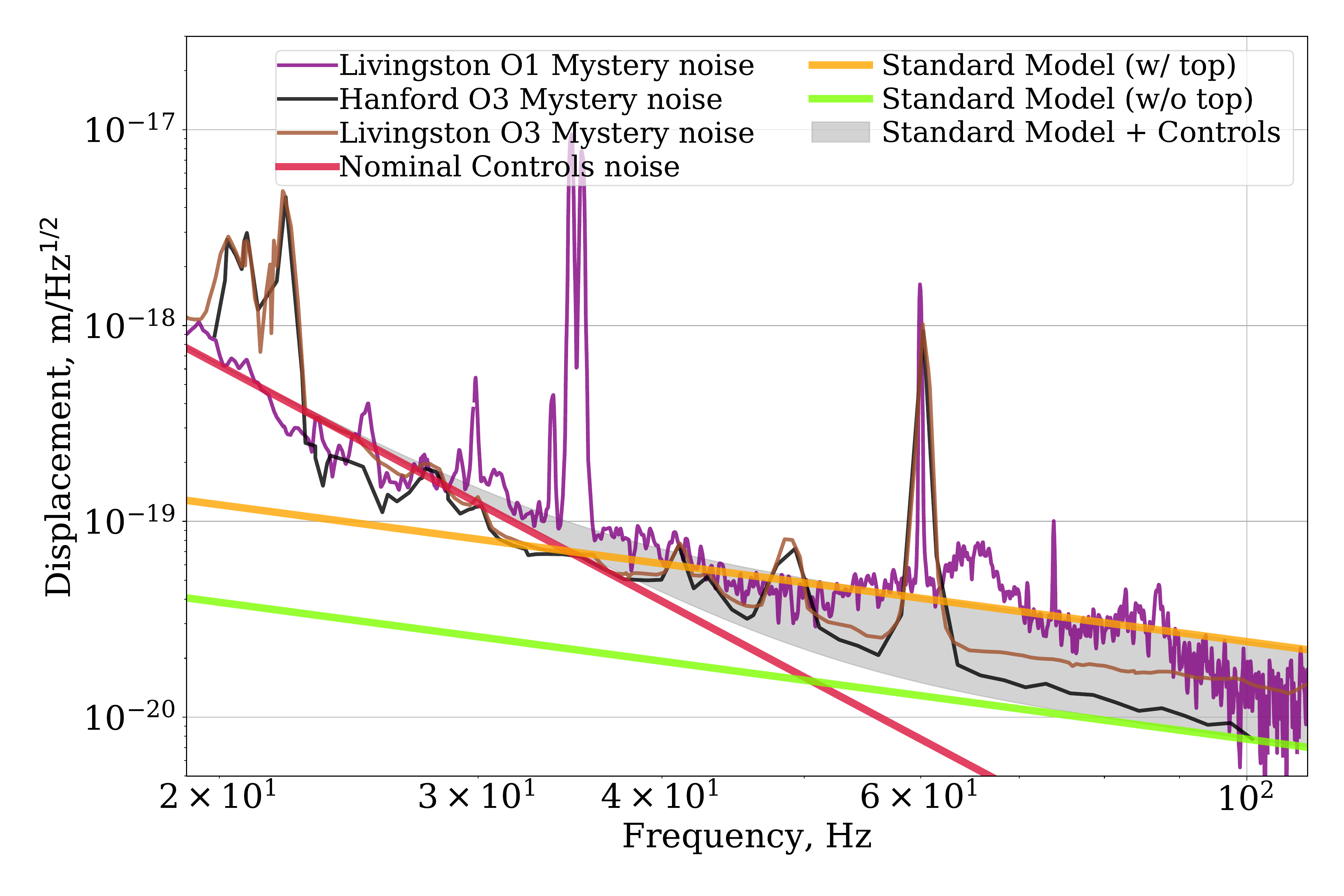}
\caption{\label{fig:mystery} Unattributed ``mystery'' noise in LIGO Livingston O1 and O3 runs, as well as Hanford O3. While the signal below 40 Hz could be attributed to Controls noise leakage $\propto f^{-4}$ (red line), the signal above 40 Hz is consistent with metric perturbations sourced by the Standard Model (SM) of Particle Physics (Equations \ref{CnC_top}-\ref{CnC_no_top}). The gray region is bound by the Controls Noise + SM contribution with and w/o top quark, as the top quark signal has ${\cal O}(1)$ uncertainty (see text for details).}
\end{figure}

While for pulsar timing observations we have \\ distance $\times$ frequency  $\gg $ speed of light, gravitational wave interferometers such as LIGO are in the opposite regime. In this regime, we can treat the long wavelength metric fluctuations quasi-statically, leading to a displacement in the interferometer arms' lengths: $\delta L \equiv L_x-L_y \simeq \frac{1}{2} L \times (h_{xx}-h_{yy})$. We can then use Einstein equation in the Lorentz gauge $\Box \bar{h}_{\mu\nu} = 2M^{-2}_{\rm P} T_{\mu\nu}$ (for trace-reversed metric perturbations: $\bar{h}_{\mu\nu} \equiv h_{\mu\nu}-\frac{1}{2}h\eta_{\mu\nu}$) and Equation (\ref{stress}) to find:      

\begin{eqnarray}
&&\langle \bar{h}_{ij}(t)\bar{h}_{mn}(t')\rangle = (\delta_{ij}\delta_{mn}+\delta_{im}\delta_{jn}+\delta_{in}\delta_{jm})\times \nonumber\\  && -\frac{\sum_ic_i m_i^5}{225 M^4_{\rm P}} \int \frac{d\omega d^3k}{(2\pi)^4} e^{i\omega(t-t')} \left[\frac{8k^4-20k^2\omega^2+15 \omega^4}{(k^2-\omega^2)^{9/2}}\right],\nonumber\\    
\end{eqnarray}
leading to the effective dimensionless strain power spectrum:
\begin{eqnarray}
&& \langle h^2_{\rm LIGO} \rangle \equiv \frac{\omega}{\pi L^2} \int dt \exp[i\omega(t-t')] \langle \delta L(t)\delta L(t')\rangle \nonumber\\
&&= -\frac{\omega \sum_ic_i m_i^5}{225 \pi M^4_{\rm P}} \int \frac{d^3k}{(2\pi)^3}\frac{8k^4-20k^2\omega^2+15 \omega^4}{(k^2-\omega^2)^{9/2}}
\nonumber \\ && = \frac{2\sum_ic_i m_i^5}{525 \pi^5 M^4_{\rm P} \times {\rm frequency}}, \label{CnC_general}
\end{eqnarray}
where $\omega({\rm rad/s}) = 2\pi \times$ frequency(Hz). Note that the $k$-integral has divergences at $k=\omega$, associated with the branch point in the integrand. Physically, this divergence is due to on-shell gravitational waves (or kernel of the $\Box$). In order to separate the gravity of vacuum stress fluctuations from ordinary gravitational waves, we can simply discard these power-law divergences (by expanding in powers of $k_{\rm min}-\omega$), which is equivalent to integrating around the branch cut in the complex plane \cite{Afshordi:2015iza}. The finite part is the last line in Equation (\ref{CnC_general}), which (fortunately) happens to be positive.    

Now, let us compute the magnitude of this effect for known particles. The main Standard Model (SM) contribution to this signal is from the most massive particles, i.e. top quark, as well as Higgs, Z, and W bosons with $m_i = (173,125,91,81)$ GeV and $c_i =(4,1,3,6)$, respectively. This yields: 
\begin{eqnarray}
&& \langle h^2_{\rm LIGO} \rangle \simeq \frac{3.70 \times 10^{-43}}{\rm frequency(Hz)} ~~{\rm (for~SM),}\label{CnC_top} \\
&& \langle h^2_{\rm LIGO} \rangle \simeq \frac{3.73 \times 10^{-44}}{\rm frequency(Hz)} ~~{\rm (for~SM~w/o~top).}
\label{CnC_no_top}
\end{eqnarray}
We see that metric perturbations computed in Equation (\ref{CnC_general}) is clearly dominated by the top quark in SM, as it is the most massive and has 4=2$\times$2 polarizations (2 for anti-top, and another 2 for spin states). However, Equation (\ref{stress}) was derived for free fields. We can estimate the relative correction to stress correlators due to interactions to be \#gluons$\times \alpha_s \sim 8 \times 0.1 \sim 1$, where $\alpha_s \sim 0.1$ is the square of QCD coupling constant at $\sim$ 200 GeV. Therefore, the exact contribution from vacuum stress fluctuations of top quark has ${\cal O}(1)$ uncertainty.     

The predictions of Equations (\ref{CnC_top}-\ref{CnC_no_top}) for the Standard Model is shown in Figure (\ref{fig:mystery}) as the yellow and green lines respectively, where {\it Displacement} noise is defined as  $ L \times \sqrt{\langle h^2_{\rm LIGO} \rangle/{\rm frequency}}$ \footnote{$L=$3994.5 m being the length of the LIGO arms \cite{Martynov:2016fzi} }. We compare this with the reported ``mystery'' noise from Livingston O1 \cite{Martynov:2016fzi}, Livingston O3 \footnote{\href{https://alog.ligo-la.caltech.edu/aLOG/index.php?callRep=48797}{https://alog.ligo-la.caltech.edu/aLOG/index.php?callRep=48797}}, and Hanford O3 \footnote{\href{https://alog.ligo-wa.caltech.edu/aLOG/index.php?callRep=50104}{https://alog.ligo-wa.caltech.edu/aLOG/index.php?callRep=50104}} which could not be attributed to any known source of noise. Quite surprisingly, Standard Model provides an excellent fit to the mystery noise for frequencies $> 40$ Hz, {\it without any free parameters}. I let the readers decide for themselves the significance of this coincidence.

The noise below 40 Hz appears to be dominated by the ``technical noise from the control systems'', which we fit by a simple power-law $(10^{-13} {\rm m/\sqrt{Hz}}) \times f({\rm Hz})^{-4}$, shown by the red line in Figure \ref{fig:mystery} \footnote{\href{https://caltechexperimentalgravity.github.io/research/intelligent-controls.html}{https://caltechexperimentalgravity.github.io/research/intelligent-controls.html}}.

Let me summarize: We have computed the spacetime metric fluctuations sourced by the quantum vacuum of the Standard Model of Particle Physics, and shown that it is consistent with the reported unattributed ``mystery'' noise in LIGO detectors, with no free parameters. If our inference is correct, it implies that there could be no weakly-coupled particle heavier than those  in standard model, radically trimming the landscape of Beyond Standard Model theories. Furthermore, it provides an irreducible noise for gravitational wave detectors that should be included in the forecasts/designs of the next generation of detectors. 

The next few years will be crucial, to see whether the ``mystery'' noise remains persistent over detector upgrades, and for different detectors (e.g., KAGRA in Japan or LIGO India), complemented by theoretical computation of stress correlators that include interacting fields.

\begin{acknowledgements}
I would like to thank Rana Adhikari, Lisa Barsotti, Thomas Dent, Gabriela Gonzalez, Denis Martynov, and Jess McIver for discussions regarding the LIGO ``mystery noise''. I am also indebted to Elliot Nelson for our original collaboration that uncovered the ``cosmological non-constant'' phenomenon.  I further thank Joao Magueijo for his comments and encouragement through this endeavour. Research at the Perimeter Institute is supported by the Government of Canada through Industry Canada, and by the Province of Ontario through the Ministry of Research and Innovation.
\end{acknowledgements}

\bibliography{mystery}
\end{document}